\documentclass[journal]{IEEEtran}

\usepackage{flushend}
\usepackage{mathrsfs}
\usepackage{color}
\usepackage{multirow}
\usepackage[flushleft]{threeparttable}

\ifCLASSINFOpdf
   \usepackage[pdftex]{graphicx}
\else
  \usepackage[dvips]{graphicx}
\fi

\usepackage{amsmath}
\usepackage{makecell}

\begin{document}

\title{Investigating Millimeter-Wave Thin-film Superconducting Resonators: A Study Using Tunnel Junction Detectors}

\author{Wenlei~Shan,~\IEEEmembership{Member,~IEEE,}
        and~Shohei~Ezaki~\IEEEmembership{}% <-this % stops a space
\thanks{Manuscript submitted March 18, 2024.}
\thanks{The work is partly supported by the Japan Society for the Promotion of Science(JSPS) KAKENHI under Grant Number 18K03708.}
\thanks{Wenlei Shan, and Shohei Ezaki are with National Astronomical Observatory of Japan(NAOJ), Osawa 2-21-1, Mitaka, 181-8588, Tokyo, Japan (e-mails: wenlei.shan@nao.ac.jp;shohei.ezaki@nao.ac.jp;).  Wenlei Shan is also with the Graduate University for Advanced Studies (SOKENDAI).}% <-this % stops a space
}

\markboth{Journal of \LaTeX\ Class Files,~Vol.~XX, No.~X, MARCH~2024}%
{Shell \MakeLowercase{\textit{et al.}}: Bare Demo of IEEEtran.cls for IEEE Journals}

\maketitle

\begin{abstract}
Investigations into the propagation characteristics, specifically loss and wave velocity, of superconducting coplanar waveguides and microstrip lines were conducted at a 2 mm wavelength. This was achieved through the measurement of on-chip half-wavelength resonators, employing superconductor-insulator-superconductor tunnel junctions as detectors. A continuous wave millimeter wave probe signal was introduced to the chip via a silicon membrane-based orthomode transducer. This setup not only facilitated the injection of the probe signal but also provided a reference path essential for differential measurements. The observed resonance frequencies aligned closely with theoretical predictions, exhibiting a discrepancy of only several percent. However, the measured losses significantly exceeded those anticipated from quasi-particle loss mechanisms, suggesting the presence of additional loss factors. Notably, the measurement results revealed that the tangential loss attributable to the dielectric layer, specifically silicon dioxide, was approximately $\rm{7\pm 2 \times 10^{-3}}$. This factor emerged as the dominant contributor to overall loss at temperatures around $\rm{4\,K}$.

\end{abstract}

\begin{IEEEkeywords}
Radio astronomy, Monolithic microwave integrated circuits, Superconducting thin-film transmission lines, Transmission loss.
\end{IEEEkeywords}

\IEEEpeerreviewmaketitle

\section{Introduction}

\IEEEPARstart{T}{hin}-film superconducting transmission lines, operating at millimeter (mm) and sub-millimeter (sub-mm) wavelengths, serve as fundamental components in the construction of superconducting detectors for astronomical observations. The fidelity of these transmission lines' actual transmission characteristics is pivotal, as it directly influences the precision of the detector's design implementation and, consequently, its sensitivity. Given their critical role within a superconducting receiver's architecture, conducting experimental evaluations of thin-film superconducting transmission lines is essential.

The experimental evaluation of thin-film superconducting transmission lines at mm/sub-mm wavelengths presents significant challenges, unlike the more straightforward approaches available in the microwave regime, where tools like network analyzers and cryogenic probe stations are readily accessible. Despite the availability of probe stations designed for either cryogenic conditions (suited for microwave frequencies) or mm/sub-mm wavelengths (for room temperature measurements) independently, integrated systems that can fully accommodate both conditions -- $\rm{4\, K}$ temperature and mm wave operation -- are still in development \cite{Russell2012,West2022}. This gap has resulted in a dearth of experimental studies leveraging on-chip resonators and detectors \cite{Javadi1992,Vayonakis2002,Gao2009,Hahnle2020,Chang2014}, with most research driven by the development of incoherent superconducting detectors, such as Microwave Kinetic Inductance Detectors (MKIDs) \cite{Gao2009,Hahnle2020} and Transition Edge Sensors (TESs) \cite{Chang2014}. These measurements typically occur at sub-Kelvin temperatures, where MKIDs and TESs are operational. Measurements at 4 K have been exclusively conducted using tunnel junctions as detectors\cite{Javadi1992,Vayonakis2002}, with loss measurement results specifically reported in \cite{Vayonakis2002}. The significant scarcity of measurement data hampers the precise design of superconducting circuits incorporating thin-film transmission lines.

\begin{figure*}[!t]
\centering
\includegraphics[width=6in,clip]{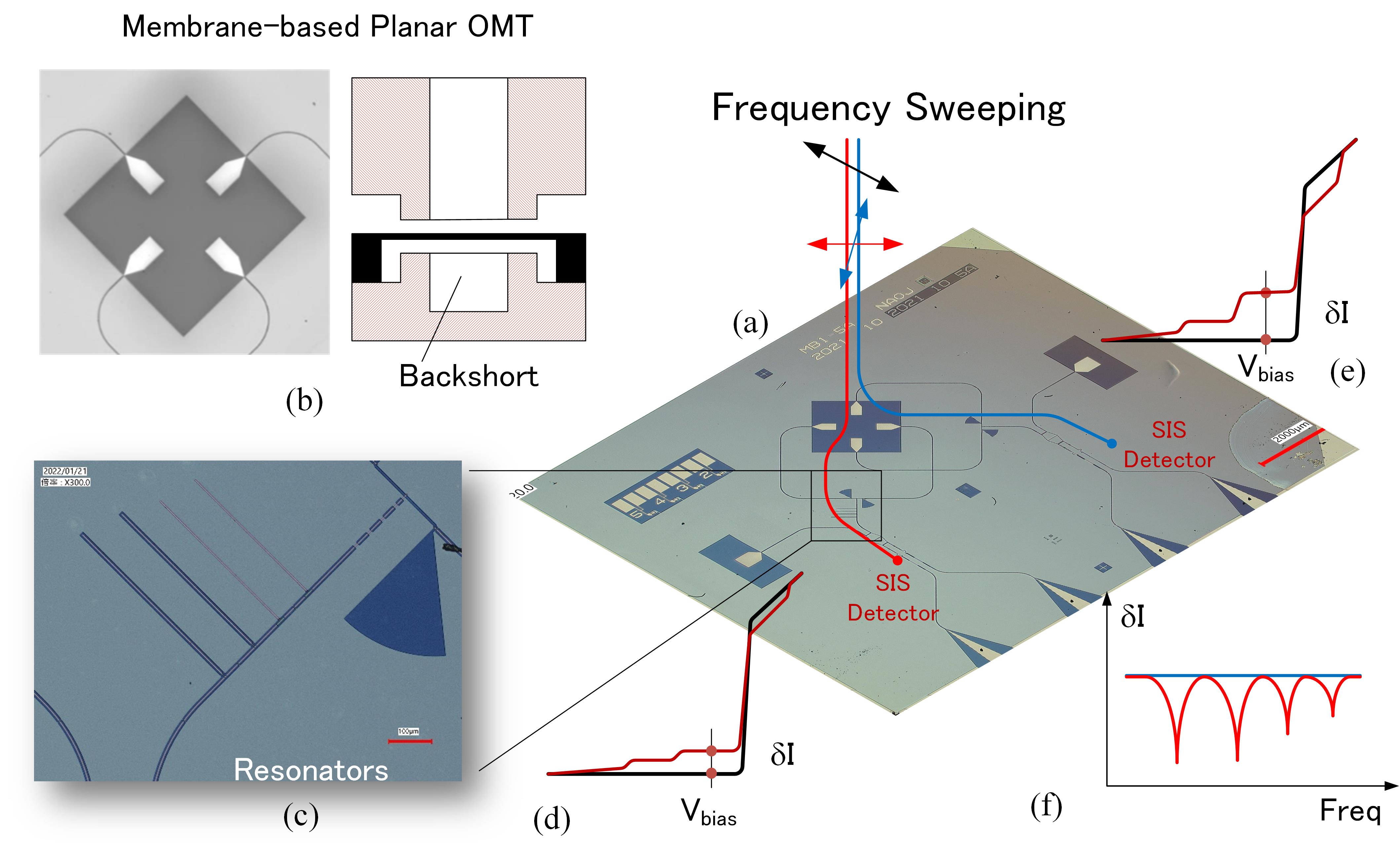}
\caption{Measurement scheme.  (a) photo of an MMIC mixer. Membrane-based planar OMT equally divides the incident signal into two streams (indicated with red and blue respectively), each of which is sensed by  niobium SIS junctions.  The red channel contains resonators, while the blue one (the reference channel) does not. (b) details of membrane-based OMT, (c) details of four niobium thin-film resonators, two with CPW architecture and two with MS one, coupled in parallel to the signal path. (d) and (e) show schematically the response of the detectors of the two channels. (f)  shows schematically the resonance curves after making corrections by using the reference channel.}
\label{FigMeasurementScheme}
\end{figure*}

We have been advancing the development of superconductor-insulator-superconductor (SIS) mixers integrated within silicon monolithic microwave integrated circuits (MMICs). This innovative approach facilitates the creation of compact, focal plane heterodyne detector arrays, crucial for broadening the field-of-view in astronomical observations at mm/sub-mm wavelengths \cite{Shan2020,Shan2019,Shan2018}. The MMIC-based mixers uniquely detect two orthogonal polarizations through dual balanced mixers, utilizing coplanar waveguide (CPW) and microstrip (MS) lines for signal and local oscillator (LO) transmission, as well as in crafting planar circuit components. Diverging from traditional SIS mixers that employ probe-like quartz substrates, our silicon-based MMIC mixers incorporate extended superconducting transmission lines, spanning several wavelengths. This design shift enables the integration of radio frequency (RF) circuit components—typically realized with bulky metal waveguide structures, such as the orthomode transducer (OMT) and hybrid coupler—directly onto the silicon chip using thin-film superconducting transmission lines. Given the lengthier transmission paths inherent to this design, the low-loss attribute of thin-film transmission lines becomes even more critical than in quartz-based SIS mixers, underscoring the necessity of thorough transmission property investigations as a cornerstone of MMIC SIS mixer development. Driven by the need for efficient diagnostic tools within this context, we have devised and measured on-chip resonators based on the mixer MMIC chips, aiming at enhancing mixer performance without compromising its integrity. This method emerges as a promising in-situ diagnostic technique, potentially pivotal in the assessment and optimization of noise performance in these sophisticated mixing devices.

\section{Method}

\subsection{Measurement Scheme}
The measurement approach and the MMIC SIS mixer utilized in this investigation are depicted in Fig.\ref{FigMeasurementScheme}. Central to Fig. \ref{FigMeasurementScheme} is the depiction of the dual-polarization, balanced SIS MMIC mixer chip. This chip is consistent with the designs previously detailed in \cite{Shan2020,Shan2019,Shan2018}, with the exception that four resonators (two MS resonators and two CPW ones) are now coupled to the signal path in one of the dual polarization channels.The polarization channel without resonators serves as a reference. Within each channel, SIS junctions function as direct detectors, capturing the power of the transmitted signal by measuring the photon-assisted tunneling current through a DC bias module. The incident signal is introduced with vertical polarization, oriented at a 45-degree angle relative to the linear polarization axes of the OMT. This arrangement ensures that each polarization channel receives an equal share of the incident power, a result of geometric symmetry. It is important to note that each polarization channel operates under a balanced mixer design. Here, the incoming signal is equally split into two paths by a 3-dB RF hybrid bridge prior to reaching the SIS detectors. The aggregate signal power in one polarization channel is determined by summing the responses from both SIS detectors. Due to this summation process, the exact balance of the RF hybrid bridge does not impact the measurement outcome.

\begin{figure}[ht]
\centering
\includegraphics[width=3.4in,clip]{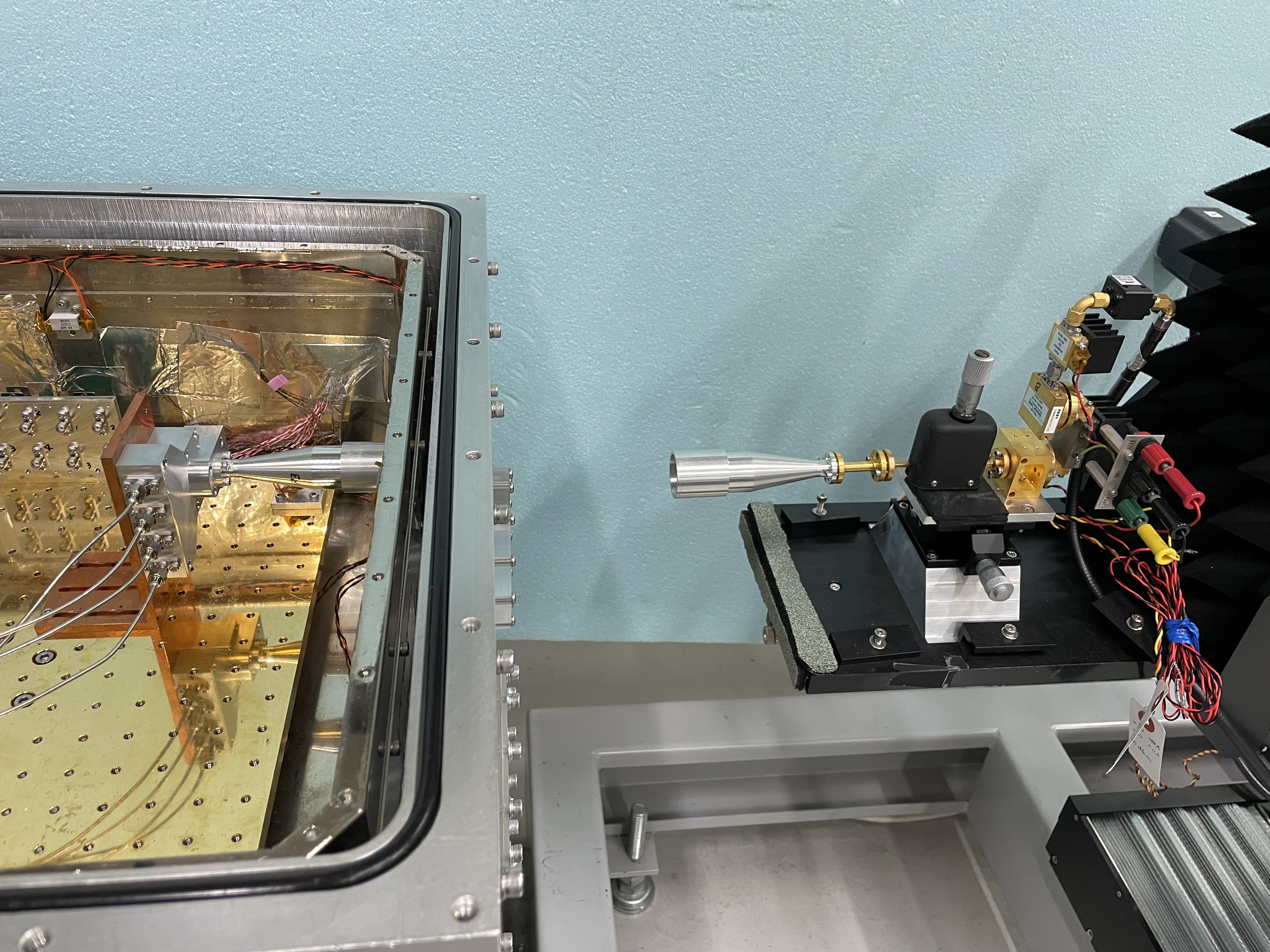}
\caption{Measurement setup. The MMIC, which is mounted in a mixer holder at the cold stage of the cryostat, detects the CW signal  through a vacuum window. The signal is generated with a CW source outside the cryostat. The feed horns of the receiver and transducer are aligned and the polarization of incident signal is 45 degree with respect to the two linear polarizations of the planar OMT integrated in the MMIC.}
\label{FigMeasurementSetupMMIC}
\end{figure}

The cornerstone of our measurement architecture is the OMT, which facilitates a reference path critical for accurate assessment. The frequency-dependent response captured in our measurements reflects the dispersion properties of the entire signal path. This encompasses not only the resonators but also the CW source and any components integrated between the source and the resonators. For resonators with a sufficiently high quality factor (Q factor), where the transmission coefficients of all ancillary components remain relatively constant within the resonator's narrow resonance frequency range, it's possible to normalize the resonance curve by using the response at frequencies immediately adjacent to the resonance as applied in high-Q resonators typically operating at deep cryogenic temperatures \cite{Gao2009,Hahnle2020}. However, our experiments encounter a challenge: the relatively low Q factors of our resonators do not meet the stringent conditions necessary for the aforementioned normalization technique. As a result, the inclusion of a reference channel—an identical signal path but devoid of resonators—becomes indispensable. This setup allows us to correct for the frequency-dependent characteristics of components other than the resonators, ensuring the accuracy of our measurements as done in \cite{Vayonakis2002}.

Figure \ref{FigMeasurementSetupMMIC} illustrates the measurement setup. The mixer block is installed inside a cryostat, achieving a minimum temperature of $\rm{3.3\, K}$ through the use of a mechanical cooler. Stage temperature monitoring is accomplished with a calibrated temperature sensor. Temperature tuning up to above the transition temperature of niobium is done by a tunable heater attached to the cold stage, which is used in the investigation of the temperature-dependence of transmission properties, aiding in the analysis of loss mechanisms within the transmission lines. Externally, the CW source is precisely aligned with the receiver's feed horn. The CW source assembly comprises a microwave signal generator, followed by an amplifier and a $\times 9$ multiplexer, facilitating resonance curve measurements across a broad frequency range of
$\rm{123-165\, GHz}$ by sweeping the signal source's frequency. To maintain the operational condition of the MMIC as a mixer, a weak magnetic field was applied perpendicularly to the MMIC from the mixer block's backside using a permanent magnet. This field, an empirical method for slightly suppressing mixer gain, enhances stability. However, We will demonstrate that this magnetic field contributes to observable losses in the transmission lines. Additionally, the incident CW signal power was verified to avoid saturating the detection response, maintaining linearity within 10\%.

\subsection{Resonator Design}

Four half-wavelength open-ended resonators, comprising two CPW and two MS structures were intentionally designed to be evenly separated in a frequency range of $\rm{123-165\,GHz}$, bounded by the capabilities of the CW source, and are capacitively coupled to the signal path. The detailed structures and the evaluated characteristics of these transmission lines are encapsulated in Table \ref{TbLineProperties}, with identical CPW and MS configurations employed within the mixer circuit itself. A significant component in the MS design is the $\rm{SiO_2}$ dielectric layer, applied via plasma-enhanced chemical vapor deposition (PECVD), as elaborated in \cite{Ezaki2020}. This layer was also purposefully deposited beneath the CPW's center strip to electrically isolate grounding bridges, effectively connecting the grounds on either side of the strip. Positioned in an area of considerable electric field strength, this layer markedly impacts the CPW's characteristic impedance and loss. For MS characteristics, a conformal transformation technique is utilized to account for the strip's thickness, on par with that of the dielectric layer \cite{Chang1977}, while the superconductor's kinetic inductance and loss are integrated through complex conductivity \cite{Kautz1978}. Calculations for CPWs, considering their multilayer composition, also employ conformal mapping \cite{Gevorgian1995}, with the complex conductivity applied analogously to the MS analysis. The precision of these analytical techniques was corroborated through numerical simulations of both CPW and MS configurations using HFSS \cite{HFSS2021}, an electromagnetic field simulator. By assigning complex surface impedance, these simulations aim to replicate the superconductor's loss and kinetic inductance, as previously demonstrated \cite{Kerr1999}. The comparison indicates that the analytical and numerical methods converge, presenting a small discrepancy of less than 5\% in characteristic impedance and wave velocity.

\begin{table}[hbt]
\caption{Structures and calculated characteristics of CPW and MS \tnote{*}}
\begin{center}
\begin{threeparttable}
\renewcommand{\TPTminimum}{\linewidth}
\makebox[\linewidth]{
\label{TbLineProperties}
\centering
\begin{tabular}{cc}
\includegraphics[width=2cm,clip]{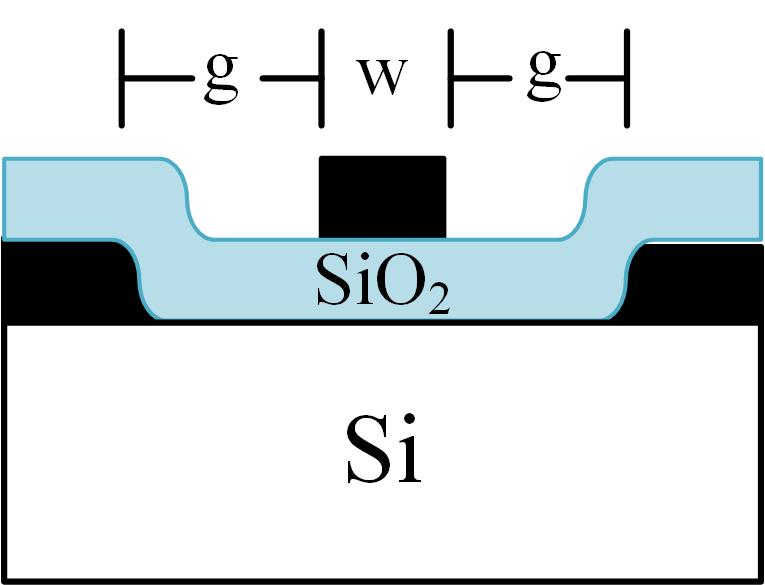}&
\includegraphics[width=2cm,clip]{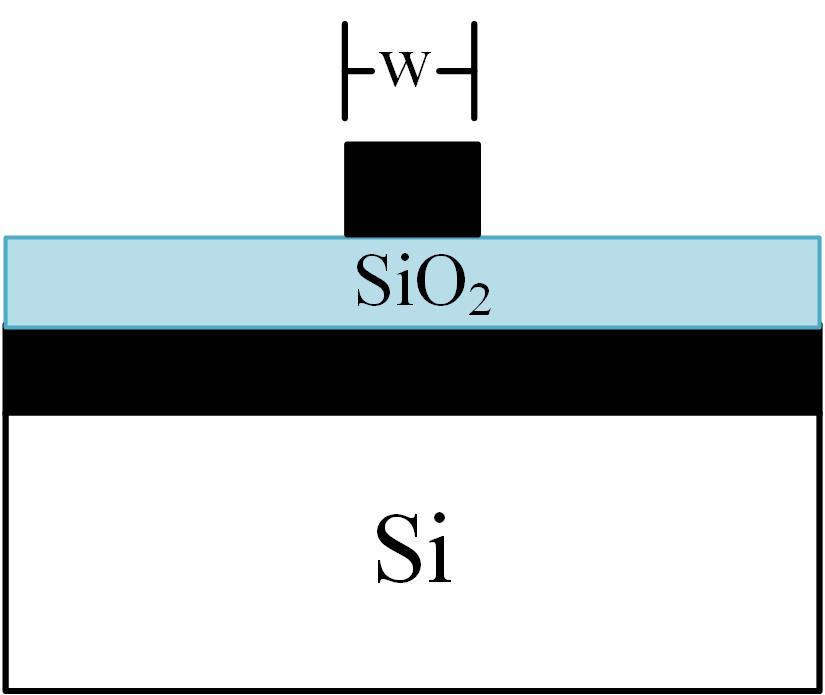}\\
\end{tabular}}
\makebox[\linewidth]{
\begin{tabular}{ccccc}
\hline
Type & w & g & $Z_0$ & $v$ \\
\hline
CPW & $3\,\mu m$ & $4\,\mu m$ & $70.2\,\Omega$  & $0.435c$ \\
MS  & $3\,\mu m$ &            & $18.4\,\Omega$  & $0.440c$ \\
\hline
\end{tabular}}
\begin{tablenotes}
\small
\item[*] The ground plane and conducting strip are fabricated from niobium films with nominal thicknesses of $\rm{300\,nm}$ and $\rm{500\,nm}$ respectively. The dielectric layer is made of $\rm{SiO_2}$ with a thickness of approximately $\rm{300\,nm}$. The wave velocity ($v$) is normalized by the light speed in free space ($\rm{c}$). $Z_0$ represents characteristic impedance. $\rm{w}$ and $\rm{g}$ are widths of the conducting strip  and the gap between the center strip and ground plane in the CPW configuration. The dielectric constants are set to be $4$ and $11.4$ for $\rm{SiO_2}$ and $\rm{Si}$ respectively.
\end{tablenotes}
\end{threeparttable}
\end{center}
\end{table}

\begin{figure}[tb]
\centering
\includegraphics[width=3.4in,clip]{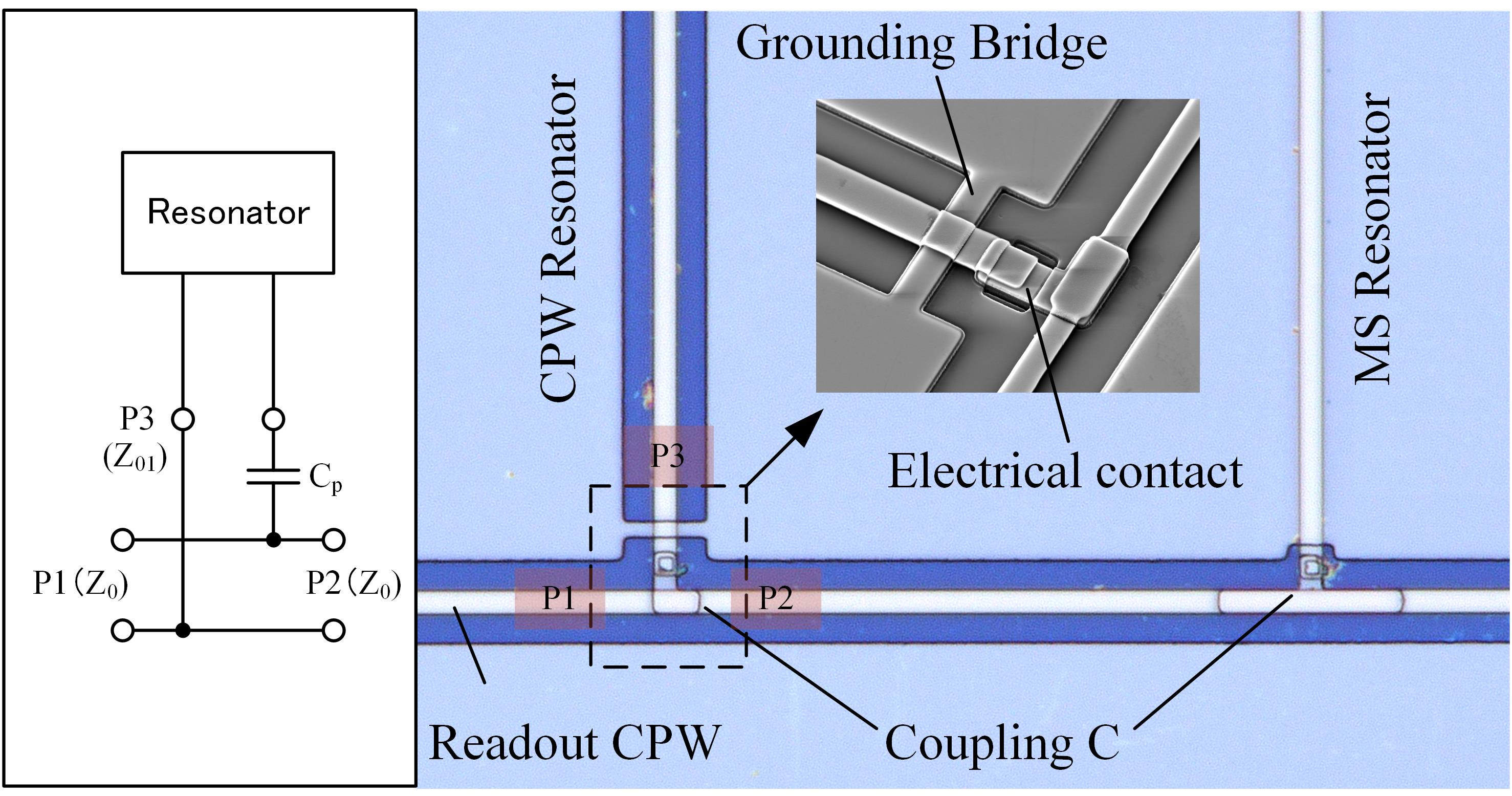}
\caption{Optical image of the coupling T-junctions of a CPW resonator and  an MS resonator. The inset shows an SEM image of the T-junction for the CPW resonator. The area of coupling capacitors is $\rm{18\,\mu m^2}$ and $\rm{72\,\mu m^2}$ respectively for CPW and MS resonators, translating to capacitances of about $\rm{3\, fF}$ and $\rm{10\,fF}$ respectively. The diagram on the left is the equivalent circuit of the T-junction used for deriving Eq. \ref{EqQc}.}
\label{FigCouplingTJunction}
\end{figure}

The half-wavelength resonators are coupled to the readout CPW line by using parallel-plate capacitors, employing the same $\rm{300\, nm}$-thick $\rm{SiO_2}$ dielectric layer as in the MS's and CPWs. An optical microscopic image illustrating the coupling T-junctions is displayed in Fig.\ref{FigCouplingTJunction}. The designed coupling quality factors for the CPW and MS resonators are approximately 120 and 30, respectively. $Q_c$ is derived using the formula $Q_c=\pi/S_{31}^2$ with the port numbering specified in the equivalent circuit depicted in Fig.\ref{FigCouplingTJunction}. Further refinement yields the expression for $Q_c$ as:
\begin{equation}\label{EqQc}
 Q_c=\frac{\pi\left(\omega^2C_p^2Z_{01}Z_0\right)}{1+\left(Z_{01}+Z_0/2\right)^2\omega^2C_p^2}\,,
\end{equation}
where $Z_{01}$ and $Z_0$ represent the characteristic impedances of the resonator and readout line, respectively, $C_p$ denotes the coupling capacitance, and $\omega$ is the angular frequency. The coupling capacitance includes the geometric capacitance of the parallel-plate capacitor, adjusted for fringe field effects, which are approximated by enlarging the capacitor's lateral dimensions by the thickness of the
$\rm{SiO_2}$ layer. HFSS simulations confirm the effectiveness of this approach for modeling fringe fields. The resonance frequency $\omega_r$ emerges from solving the nonlinear equation:
\begin{equation}\label{EqResFreq1}
 \tan\beta L+Z_{01}\omega_r C_p=0\, .
\end{equation}
yielding a condition of virtual grounding across the readout line at the input at $\omega_r$. This leads to a reformulated resonance frequency equation:
\begin{equation}\label{EqResFreq2}
 \tan[\pi(\bar\omega_r-1)]=-\alpha\bar\omega_r \, ,
\end{equation}
with $\alpha=Z_{01}\omega_0 C_p$, $\bar\omega_r=\omega_r/\omega_0$ and $\omega_0=\pi\nu/L$ representing the base resonance frequency excluding coupling capacitance effects. For sufficiently high $Q_c$, the equation simplifies, resulting in $\bar\omega_r\simeq1-Z_{01}\omega_0 C_p$.

\subsection{Data Reduction}

\begin{figure}[tb]
\centering
\includegraphics[width=3.4in,clip]{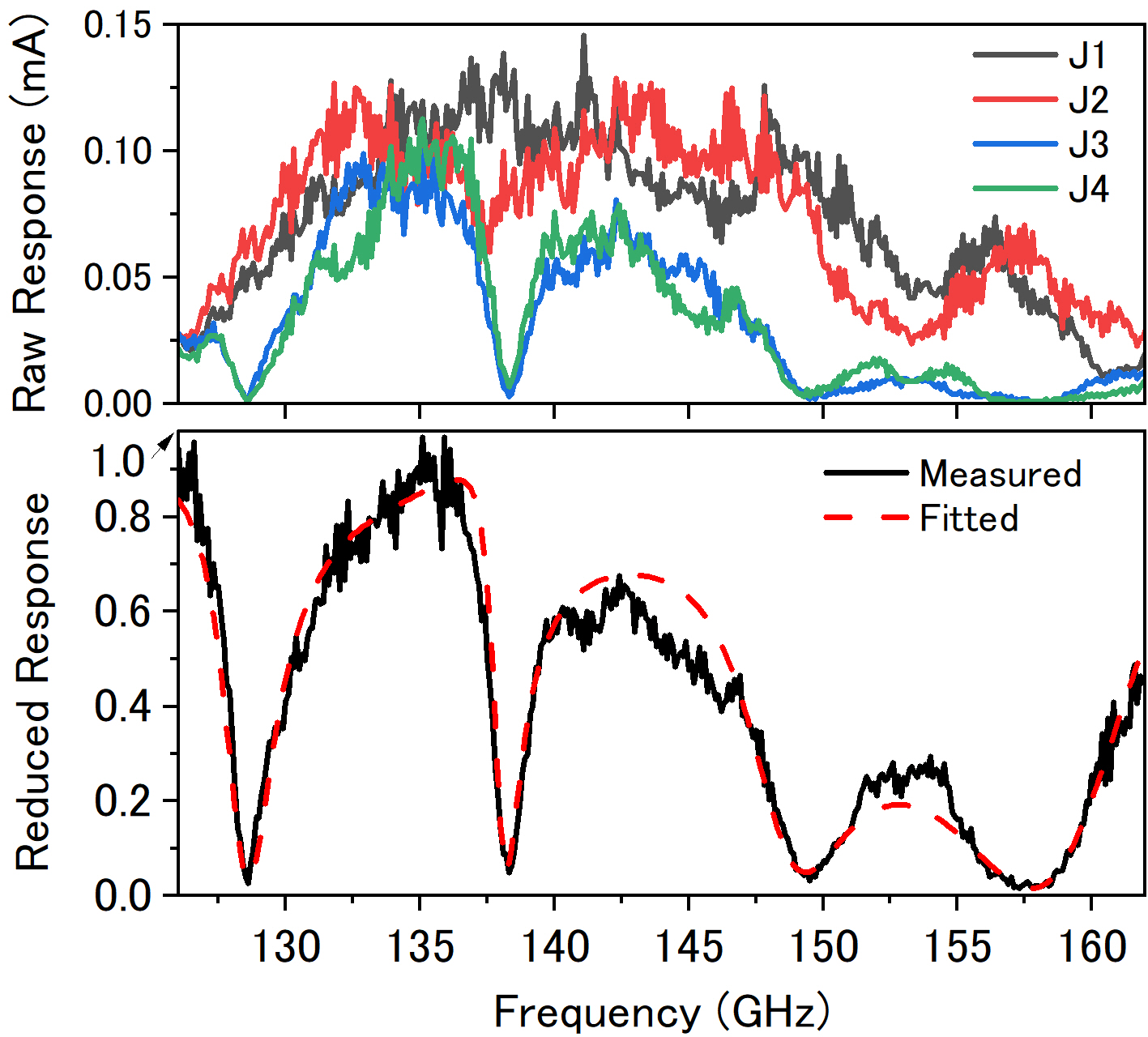}
\caption{Measured raw resonances of four detectors of an MMIC (upper panel) and the reduced resonance compared with fitting result (lower panel).}
\label{FigFittingCurves}
\end{figure}
%fitting program: My Program/Tools/ResonatorFitting/

The direct detection responses from the SIS junctions were recorded in increments of $\rm{0.1\,GHz}$. Owing to the balanced mixing architecture the MMIC mixer chip yields four output terminals. To delineate the reduced resonance curves, outputs corresponding to a single polarization are initially aggregated to capture the comprehensive response for that channel. Subsequently, this cumulative response is normalized by dividing it by the output from the reference polarization channel. Fig \ref{FigFittingCurves} presents a representative example of both the raw detection responses and the processed (normalized) response measured at a temperature of $\rm{3.3\,K}$.

\begin{table}[bt]
\begin{center}
\begin{threeparttable}
\caption{Comparison between fitted and calculated resonator parameters. \tnote{*}}
\label{TbResonatorProperties}
\centering
\begin{tabular}{ccccccc}
\hline
ID & \makecell{$L$ \\ ($\mu m$)}  & \makecell{$f_{0,clt}$\\(GHz)} & \makecell{$f_{0,fit}$\\(GHz)} & $Q_{c,clt}$ & $Q_{c,fit}$ & $Q_{0,fit}$\\
\hline
CPW1 & \rm{464} & $\rm{133.8}$ & $\rm{128.6\pm 0.1}$ & $\rm{124}$  & $\rm{97\pm1.5}$ & $\rm{228\pm10}$\\
CPW2 & \rm{433} & $\rm{142.9}$ & $\rm{138.3\pm 0.1}$ & $\rm{110}$  & $\rm{142\pm0.4}$ & $\rm{393\pm20}$\\
MS1 & \rm{391} & $\rm{158.7}$ & $\rm{149.5\pm 0.1}$ & $\rm{30}$  & $\rm{23.4\pm0.4}$ & $\rm{99\pm2}$\\
MS2 & \rm{368} & $\rm{168.0}$ & $\rm{157.9\pm 0.1}$ & $\rm{27}$  & $\rm{20\pm0.2}$ & $\rm{119\pm3}$\\
\hline
\end{tabular}
\begin{tablenotes}
\small
\item[*] $L$ denotes the length of a resonator. $f_0$ is the resonance frequency. The subscript $clt$ and $fit$ mean calculated and fitted values respectively. The measurement was done at an ambient temperature of $\rm{3.3\, K}$.
\end{tablenotes}
\end{threeparttable}
\end{center}
\end{table}
%photomask IMB1-M5a-Res-check-v1.dwg
%Device ID: IMB5A-02 B2 (20220106)
%Designed values: Coupling-CPW-MSE7-to-IMB1.xlsx
%Fitting values: 20220106\T-Dependance.opj
%Calculation: \IMB1-M5 Resonator\Coupling-MS(CPW)-MSE7-to-IMB1.xlsx

Notably, the resonance curves of adjacent resonators, especially the two MS resonators, exhibit significant overlap, necessitating a model that accounts for inter-resonator crosstalk. The collective transmission profile is approximated using:
 \begin{equation}\label{EqCoupledModel}
  T=\prod_{i=1}^{4}{T_i\left(\omega_{ri}, Q_{ti}, Q_{0i}\right)}\,,
\end{equation}
with each $T_i$ defined as:
\begin{equation}\label{EqLorentz}
 T_i=\left|1-\frac{Q_{ti}/Q_{ci}}{1+j2Q_{ti}\left(\omega-\omega_{ri}\right)/\omega_{ri}}\right|^2\,.
\end{equation}
Here, $Q_0$ represents the intrinsic Q factor, calculated as $Q_0=\left(Q_t^{-1}-Q_c^{-1}\right)^{-1}$. The fitting process extracts the $Q_{c}$, total Q factor ($Q_{t}$), and resonance angular frequency $\omega_{r}$ for each resonator indexed by $i$, involving twelve variables across all four resonators. This simultaneous fitting, utilizing the Levenberg-Marquardt Method \cite{Press1992}, aligns closely with the observed data as shown in Fig.\ref{FigFittingCurves}. This fitting model, particularly effective when resonators exhibit wide overlapping at elevated temperatures and lower Q factors, demonstrates its robustness in contrast to individual resonance curve fitting, which fails under such conditions.

\section{Result and Discussion}

\subsection{Resonator Characterization}
Table \ref{TbResonatorProperties} compares the designed parameters against those derived from the resonance curves illustrated in Fig. \ref{FigFittingCurves}. This comparison reveals that the center frequencies for CPW and MS resonators were overestimated by approximately 4\% and 6\%, respectively, according to our analytical methods. The cause of this variance remains uncertain; however, an empirical adjustment, applying a correction factor of $\rm{0.95}$ to the calculated wave velocities, significantly narrows this gap, achieve about 1\% consistence between measurement and calculation of the resonance frequencies of both MS and CPW resonators.

Although the resonance frequencies were reasonably well predicted, the measured intrinsic quality factors exhibited significant variability. Aggregating the fitting results from 19 CPW resonators and 12 MS resonators measured at $\rm{3.3\,K}$, we determined $Q_0$ values of $\rm{300\pm100}$ for CPW resonators and $100\pm 40$ for MS resonators. This spread in $Q_0$ values is notably larger than the fitting algorithm's uncertainty, which Table \ref{TbResonatorProperties} indicates is less than 10\%. This discrepancy suggests that factors related to the fabrication process introduce a significant degree of uncertainty into $Q_0$, though the specific aspects of fabrication that affect loss remain unclear. For subsequent analyses, we selected resonators exhibiting higher $Q_0$ values to compare with theoretical predictions. This choice is predicated on the assumption that these resonators are minimally impacted by fabrication-related reasons, allowing for a more focused comparison with the theoretical model.

\begin{figure}[tb]
\centering
\includegraphics[width=3.4in,clip]{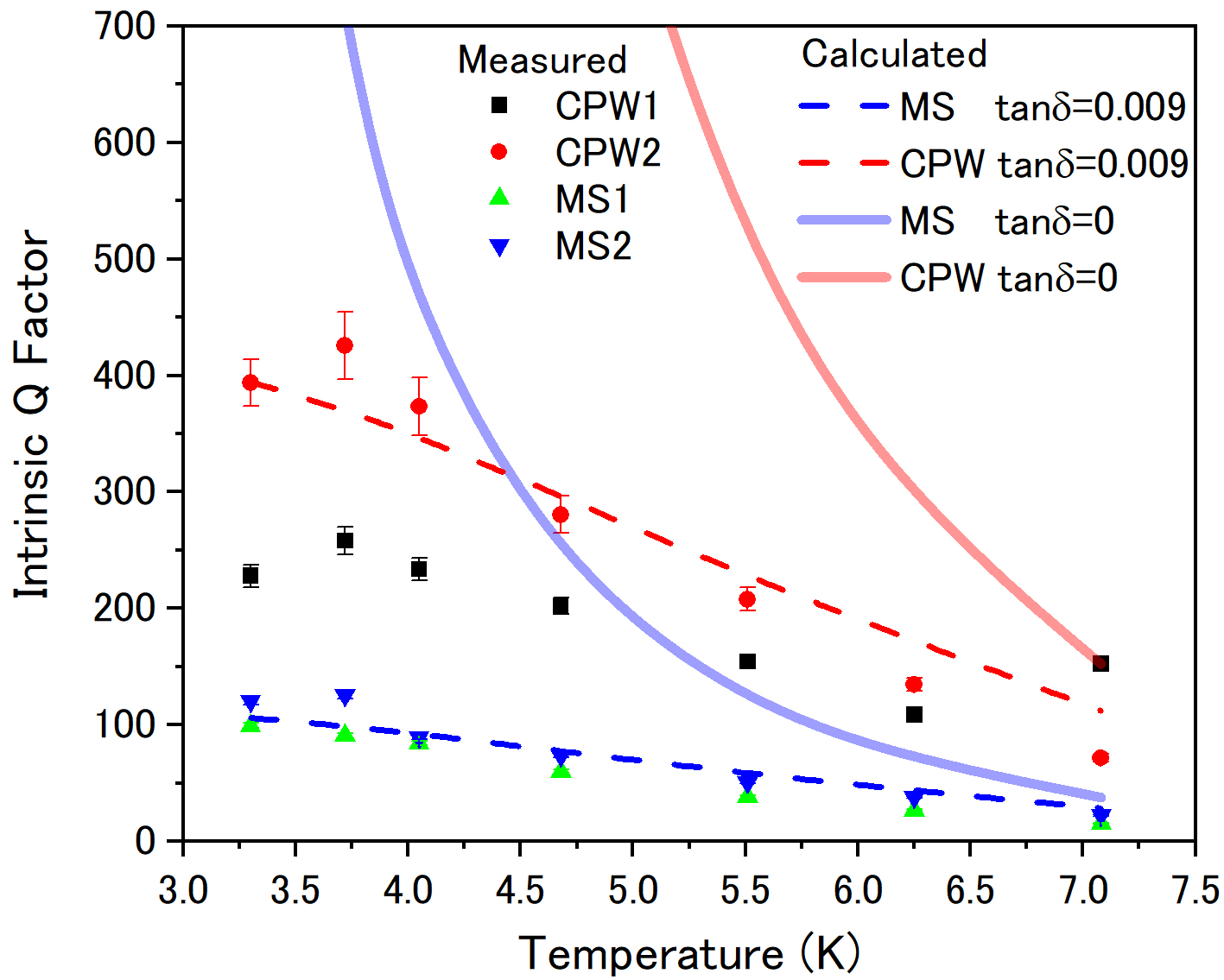}
\caption{Measured $Q_0$ versus ambient temperature and comparison with simulated $Q_0$ under two conditions: one with loss-free $\rm{SiO_2}$ layer and the other with a tangential loss of $\rm{9\times10^{-3}}$ for that layer.  }
\label{FigLossModeling}
\end{figure}
%Device IMB5A-02 B2
%Before V5 the Q factors were retrieved by using minimum transmission method (20221222/Resonance1223.opj). Since V6 Q factors were retrieved by using fitting (12 variables) method. Fitting program: /Tools/ResonatorFitting/ResonatorFitting4RV2
%Plot data /20221222/Resonance1222.opj

The unexpectedly high loss observed in our study is primarily ascribed to the conductive loss of niobium films and the dielectric loss of $\rm{SiO_2}$, given that radiation loss for CPW with $4\,\mu m/3\,\mu m/4\,\mu m$ configuration is estimated based on the method in \cite{rutledge1983integrated} to be negligible at this frequency, with a Q factor of about $\rm{7\times10^3}$, and MS resonators inherently avoid radiation loss due to their enclosed structure. To discern the contributions of quasi-particle and dielectric loss, we examined the temperature dependence of the Q factors, as these losses respond differently to changes in physical temperature. The MMIC device used to obtain the data in Fig. 4 underwent measurements across various ambient temperatures, comparing the measured intrinsic $Q_0$ against simulations as a function of ambient temperature, as shown in Fig. \ref{FigLossModeling}. A leveling off of the measured $Q_0$'s of both MS and CPW resonators at lower temperatures strongly suggests that dielectric loss, which unlikely varies significantly within the measured temperature range, is the predominant factor.

Theoretical loss calculations for CPWs and MSs were conducted using the conformal mapping methods previously outlined augmented to include various loss mechanisms. Conductive loss is assumed as resulting from thermally excited quasi-particles, which is calculated by using Mattis-Bardeen theory\cite{mattis1958theory}. Regarding dielectric losses, consideration was given solely to the $\rm{SiO_2}$ layer, as losses in the silicon substrate (made from float-zone silicon with a room temperature resistivity $\rm{>5\,k\Omega cm}$) are deemed negligible in comparison.

Two scenarios for the $\rm{SiO_2}$ layer were explored: a no-loss assumption focusing all losses on quasi-particles, and a second scenario incorporating a tangential loss of $\rm{0.9\times{10}^{-3}}$, which closely matched empirical observations. This alignment between simulated and actual $Q_0$ values for both MS resonators and the CPW2 resonator underlines the accurate integration of dielectric losses into the models for CPW and MS, each with a distinct filling factor for $\rm{SiO_2}$, and supports the predominance of dielectric loss at low temperatures. It's important to note two factors: first, the CPW1 resonator, which exhibited a lower Q factor than the first, was excluded from this comparative analysis due to potential fabrication discrepancies; second, the data in Fig.\ref{FigLossModeling} were measured with a magnetic field applied perpendicular to the MMIC with a field of about $\rm{60\,Guass}$. This field was found to result in a minor yet noticeable adjustment to the tangential loss of $\rm{SiO_2}$.

\subsection{Magnetic Field Influences}

\begin{figure}[tb]
\centering
\includegraphics[width=3.4in,clip]{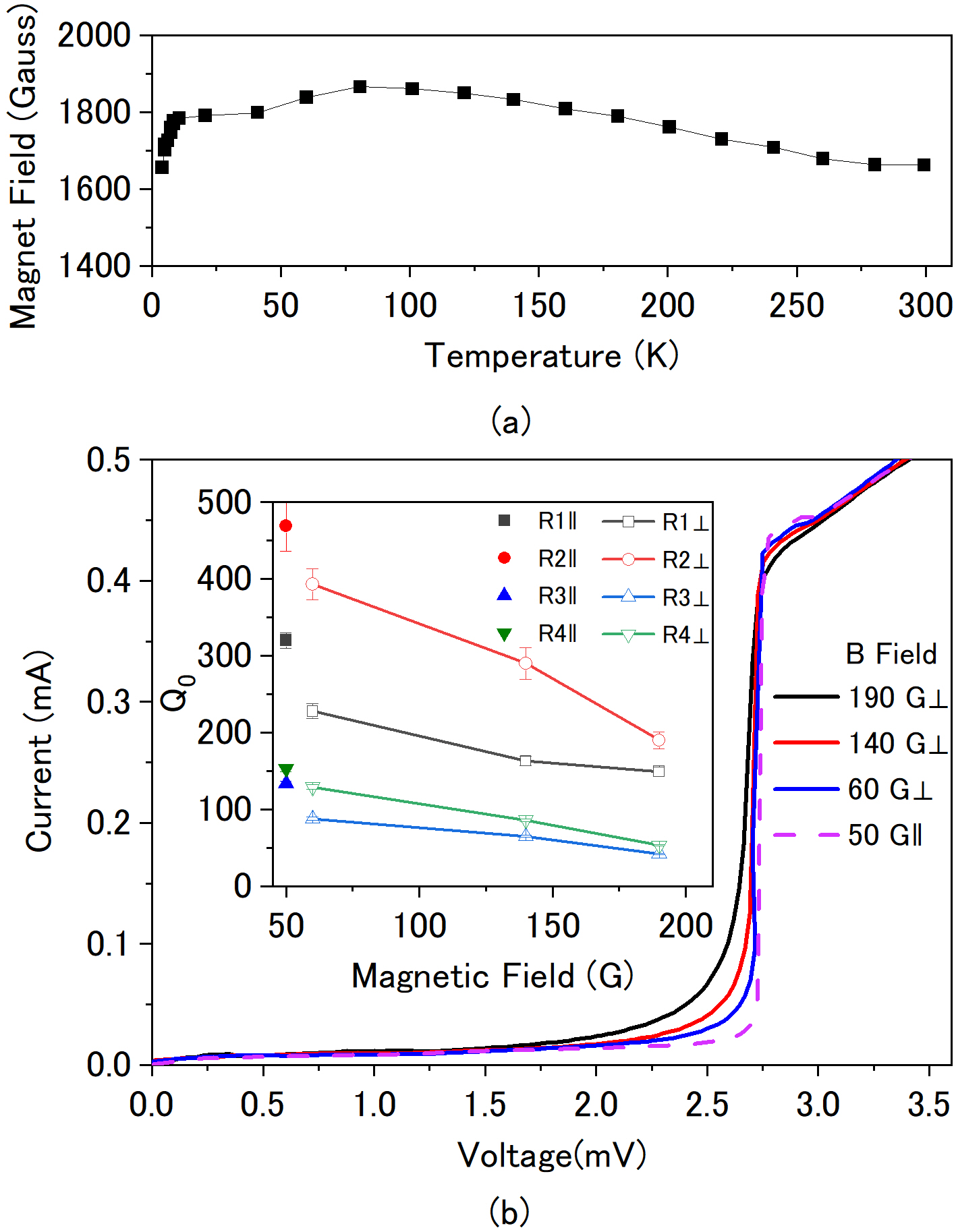}
\caption{(a) Dependence of the magnetic field of a permanent magnet on ambient temperature from room temperature to $\rm{4\,K}$. (b) Rounding of I-V curves of an SIS junction at the voltage gap caused by the magnetic field and the dependence of the resonators' Q factors on magnetic field. The same device used in Fig. \ref{FigFittingCurves} and Table \ref{TbResonatorProperties} was measured at $\rm{3.3\,K}$. }
\label{FigMagnetic}
\end{figure}

A magnetic field perpendicular to the MMIC was empirically applied to mitigate high dynamic resistance of the SIS mixer. It became essential to assess this magnetic field's influence on transmission loss, especially given the relatively lengthy transmission lines within the MMIC SIS mixer. This investigation involved positioning permanent neodymium magnets at distances of $\rm{5\, mm}$ (for a perpendicular field) or $\rm{25\, mm}$ (for a parallel field) from the MMIC chip, external to the mixer block. Due to the mixer block's design constraints, only a weak parallel magnetic field of up to approximately $\rm{50\,Guass}$ could be applied, contrasting with the broader range achievable in perpendicular field strength (up to about $\rm{200\,Gauss}$), dictated by the magnets' proximity to the chip. Magnetic field strength at the MMIC's location was gauged with a gaussmeter at room temperature. Furthermore, a cooling test from room temperature to $\rm{4\,K}$ quantified the magnetic field's temperature dependency near the magnet's surface with results shown in Fig.\ref{FigMagnetic} (a), revealing an increase in magnetic strength up to about $\rm{80\,K}$ by a factor of $\rm{1.13}$, followed by a decrease upon further cooling, ultimately aligning closely with the room temperature measurement at $\rm{4\,K}$.

According to the results shown in Fig.\ref{FigMagnetic} (b), the resonators, subjected to various magnetic field strengths adjusted by employing magnets of differing sizes, exhibited a clear trend: loss rose with increasing magnetic field strength under a perpendicularly applied field. Notably, the resonators' Q factor was larger under a parallel field than under a perpendicular field of equivalent strength. This distinction suggests the vortex motion loss's pivotal role under a perpendicular magnetic field. The I-V curve's rounding at the gap voltage onset, proportional to the magnetic field's strength, is likely attributed to the magnetic field-induced de-pairing effect\cite{Tinkham2004}. The reduction in niobium's energy gap potentially augments the quasi-particle count at finite ambient temperatures. However, this effect indirectly influences loss, as the density of state's broadening contributes to loss in an opposite way\cite{Kwon2018}. To avoid this complexity we focus on the parallel field case where no significant de-pairing phenomenon was observed and hence the influence of de-pairing to the transmission loss is negligibly small due to the weakness of the field. Given that the Q factors measured under a weakly applied parallel field are unaffected by vortex motion loss and quasi-particle loss is relatively unimportant at $\rm{3.3\,K}$ as shown in Fig.\ref{FigLossModeling}, the residual loss—presumed to originate from the $\rm{SiO_2}$ layer—closely approximates the total measured loss ($1/Q_t$) of MS resonators, estimated to be $\rm{7\pm2\times 10^{-3}}$.

\subsection{Discussion}

The dielectric loss associated with the amorphous $\rm{SiO_2}$ layer implied in our study, significantly exceeding the loss of crystalline quartz material, approximately aligns with previously published data under similar conditions. A tangential losses of $\rm{5.3\times 10^{-3}}$ for sputtered $\rm{SiO_2}$ at frequencies between $\rm{75}-\rm{100\,GHz}$ and at a temperature of $\rm{4.2\,K}$ were reported by Vayonakis et al. \cite{Vayonakis2002}, and losses ranging from
$\rm{0.5-2\times 10^{-3}}$ at approximately $\rm{100\,GHz}$ and $\rm{30\,mK}$ were observed by Gao et al. \cite{Gao2009}, with both sets of measurements conducted using microstrip resonators. The distinct temperature dependence of the dielectric loss observed at $30\,mK$ suggests the nature of two-level system (TLS) loss\cite{Gao2009}.

To determine if TLS loss contributes to the dielectric loss in our experiments, two approaches are proposed: measurement at millikelvin temperature ranges as performed by Gao et al.\cite{Gao2009}, and applying higher signal power to potentially suppress loss through stronger electric field. Unfortunately, limitations in our resources preclude millikelvin-range measurements, and higher power applications saturate the SIS detectors. Our measurement was probably conducted in the weak field limit where TLS loss saturated, as no power dependence was observed. To obtain a solid evidence of TLS noise, establishing a network analyzer-based measurement approach, allowing for higher power measurements without detector saturation, is identified as a necessary step for future studies.

The losses measured in our CPWs and MSs are markedly above those anticipated from calculations based on ideal materials and conditions. These transmission losses' impact on SIS mixer performance varies with the mixer's design. Traditional SIS mixers on quartz substrates utilize shot superconducting thin-film transmission lines primarily for impedance matching, minimizing the impact of transmission loss on receiver noise due to their brief electrical length. However, in designs employing longer transmission lines, such as the MMIC mixer detailed in this work, transmission loss becomes a critical factor. The estimated overall transmission line loss of $\sim\rm{0.6\,dB}$ based on above results—encompassing approximately four wavelengths of CPW and one wavelength of MS—translates to a noise temperature increase of about 15\%. Achieving sensitivity comparable to conventional quartz SIS mixers necessitates a cautious approach to superconducting transmission line use, particularly minimizing amorphous dielectric layers and avoiding vertical magnetic fields.

%\begin{figure}[tb]
%\centering
%\includegraphics[width=3.4in,clip]{Figures/Fig_MagnetVsT.jpg}
%\caption{Dependence of the magnetic field of a permanent magnet on ambient temperature from room temperature to $\rm{4\,K}$.}
%\label{FigMagnetVsT}
%\end{figure}
%/20220209/RF_Trans0209.opj

%Device IMB5A-02 B2
%/20220209/RF_Trans0209.opj

\section{Conclusion}

Superconducting MS and CPW resonators are characterized at millimeter-wave frequencies utilizing SIS junctions for direct detectors and an on-chip OMT to provide a reference channel. This method proves particularly beneficial in the development of MMIC SIS mixers, offering an in-situ measurement technique that enhances the understanding and optimization of these devices. The experimental findings underscore the significant impact of the dielectric loss within the $\rm{SiO_2}$ layer on the total loss encountered by superconducting transmission lines, with the extent of this influence varying according to the layer's filling factor. Moreover, the application of a magnetic field perpendicular to the superconducting strips was found to induce notable losses, likely due to vortex motion. This insight into the behavior of superconducting resonators under various conditions is crucial for refining MMIC SIS mixer designs, highlighting the need to carefully manage dielectric losses and magnetic field effects to maintain optimal performance.

%\section*{Acknowledgment}

\ifCLASSOPTIONcaptionsoff
  \newpage
\fi

\bibliographystyle{IEEEtran}

\bibliography{OnChipResonatorBibB}

\end{document}